# Biomolecular Filters for Improved Separation of Output Signals in Enzyme Logic Systems Applied to Biomedical Analysis

Jan Halámek,[a] Jian Zhou,[a] Lenka Halámková,[a,c] Vera Bocharova,[a]
Vladimir Privman,[b] Joseph Wang[d]* and Evgeny Katz[a]*

[a] Department of Chemistry and Biomolecular Science,  [b] Department of Physics,
[c] Department of Biology, Clarkson University, Potsdam, NY 13699, USA
[d] Department of NanoEngineering, University of California at San Diego, La Jolla, CA 92093, USA

## ABSTRACT

Biomolecular logic systems processing biochemical input signals and producing "digital" outputs in the form of YES/NO were developed for analysis of physiological conditions characteristic of liver injury, soft tissue injury and abdominal trauma. Injury biomarkers were used as input signals for activating the logic systems. Their normal physiological concentrations were defined as logic-**0** level, while their pathologically elevated concentrations were defined as logic-**1** values. Since the input concentrations applied as logic **0** and **1** values were not sufficiently different, the output signals being at low and high values (**0**, **1** outputs) were separated with a short gap making their discrimination difficult. Coupled enzymatic reactions functioning as a biomolecular signal processing system with a built-in filter property were developed. The filter process involves a partial back-conversion of the optical-output-signal-yielding product, but only at its low concentrations, thus allowing the proper discrimination between **0** and **1** output values.

## INTRODUCTION

Chemical[1-5] and biochemical[6-10] systems mimicking Boolean logic gates and their networks have recently been designed for novel computational and signal processing applications. Particularly rapid development has been achieved for biomolecular systems including those based on proteins/enzymes,[10,11] DNA,[9,12] RNA[13] and whole cells.[14,15] Specifically, it has been realized that operation of biomolecular computing systems in biochemical and particularly biological environments[16] enables design of novel biosensors capable to multiplex and logically process many biochemical signals in the binary format **0** and **1**, with the information processing steps carried out by chemical systems rather than electronics.[17] Several biomolecular logic systems have been designed for biomedical/diagnostic applications aiming at the analysis of biomarkers characteristic of various diseases[18-20] or injuries.[21-25]

A major difference between such systems operating as multi-signal processing biosensors and those designed for (bio)chemical computing is in the definition of **0** and **1** levels of input signals. For chemical computing, logic-**0** can be selected as the absence of a reacting species and logic-**1** defined at a conveniently high concentration of that species. In the systems designed for the logic analysis of biomedical conditions, logic-**0** and **1** should correspond instead to normal physiological and pathophysiological conditions, respectively. In some digitally operating biosensor systems the answer YES/NO is easily achieved due to a large difference in the biomarker concentrations (e.g., in the pregnancy tests). However, in many cases the difference between **0** and **1** inputs is not very large, potentially resulting in a small gap separating the output signals. This makes differentiation of the YES/NO answers of the binary biosensing systems difficult, unless careful optimization is performed.[26]

This problem can be partially solved by optimization of the readout time, getting the signals at the time when they are substantially different due to different kinetics in the reacting processes.[21,23] However, if chemical output signals are to be used for triggering chemical actuators, for example a drug releasing membrane,[27] this approach cannot be utilized because the product concentrations for the YES/NO results could reach the same levels at different reaction times. The problem of the output signal discrimination can be solved by application of chemical-filter component processes converting convex response function characteristic of most chemical processes, to sigmoidal.[28] These novel chemical filter systems were recently designed and optimized as standalone elements of logic networks.[29,30] The present Letter describes the first application of a filter system for enabling the desired binary, here **AND**, function, and improving the fidelity of signal conversion in biomedical analytical systems exemplified by logic systems for the analysis of liver injury (LI), soft tissue injury (STI) and abdominal trauma (ABT), which were recently designed[23] and then tested for robust operation in human serum solutions.[21]

**EXPERIMENTAL SECTION**

*Chemicals and reagents used:* For the LI system, alanine transaminase (ALT) from porcine heart (E.C. 2.6.1.2), lactate dehydrogenase (LDH) from porcine heart (E.C. 1.1.1.27), glucose-6-phosphate dehydrogenase (G6PDH) from *Leuconostoc mesenteroides* (E.C. 1.1.1.49), β-nicotinamide adenine dinucleotide reduced dipotassium salt (NADH), L-alanine (Ala), D-glucose-6 phosphate (Glc6P), α-ketoglutaric acid (α-KTG), *tris*(hydroxymethyl)-aminomethane (Tris-buffer), and other standard inorganic/organic reactants were purchased from Sigma-Aldrich and used as supplied. Ultrapure water (18.2 MΩ·cm) from NANOpure Diamond (Barnstead) source was used in all of the experiments. Composition and results for logic systems for the analysis of STI and ABT operated in buffer solutions, as well as for the LI system operated in human serum, are presented in Supporting Information, were the chemicals used for the STI and ABT systems are specified.

*Instrumentation and measurements:* A Shimadzu UV-2450 UV-Vis spectrophotometer, with a TCC-240A temperature-controlled holder and 1 mL poly(methyl methacrylate) (PMMA) cuvettes, were used for all optical measurements. All optical measurements were performed in temperature-controlled cuvettes at $37.0 \pm 0.2$°C mimicking physiological conditions, and all reagents were incubated at this temperature prior to measurements. In order to facilitate the signal-conversion quality analysis assessing the analytical variance present (as described later), four repeated experiments were performed for each of the four input combinations, with and without the signal filtering process.

*Composition and operation of systems for the analysis of injury:* The LI analyzing system was realized in the Tris-buffer, 100 mM, pH 7.4, used as a background solution. Ala (200 mM), α-KTG (10 mM) and NADH (300 μM) were dissolved in this solution to perform the **AND** logic operation. G6PDH (10 U·mL$^{-1}$) and Glc6P (4 mM) were added to the solution to perform the filter operation. Logic **0** and **1** levels of ALT (0.02 and 2 U·mL$^{-1}$) and LDH (0.15 and 1 U·mL$^{-1}$) input signals were selected in order to mimic meaningful circulating levels of these biomarkers under normal and pathophysiological conditions, respectively. The output signal corresponding to the decreasing concentration of NADH was measured optically at λ = 340 nm and defined as the absolute value of the absorbance change.

**RESULTS AND DISCUSSION**

The two enzyme inputs, ALT and LDH, are biomarkers characteristic of liver injury (LI).[31] It should be noted that each of them is not specific enough to indicate LI, however their simultaneous increase in concentration, from normal to pathophysiological levels, provides an unambiguous evidence of LI.[21,23] Based on the sequence of the biochemical reactions (Scheme 1A), the final result — NADH oxidation followed by absorbance change (Figure 1A) — should happen only upon cooperative work of both enzyme-biomarkers. However, it should be remembered that logic-**0** values in their present definition are not the absence of the enzymes, but rather their presence at normal physiological levels. Therefore, the reactions proceed not only at the **1**,**1** combination of the input signals, but also, albeit at slower rates, when the inputs are supplied at the **0**,**0**, **0**,**1** and **1**,**0** combinations.

When the readout time interval is limited to 50-200 sec, the absorbance decrease is sufficiently larger for the **1**,**1** input combination (implying output signal **1**) as compared to three other input combinations that yield smaller absorbance changes (output signals **0**) (Figure 1A). However, when the experiment continues to times larger than 200 sec, which is important for certain recently investigated actuation applications,[32] the absorbance decrease for the **1**,**0** input combination becomes comparable with one for **1**,**1** inputs (Figure 1A). For even larger times, the results for the **1**,**0** and **1**,**1** input combinations will not be distinguishable, and the logic function can no longer be regarded as an **AND** gate (Figure 1B, Inset: black bars). In order to increase the gap separating **0** and **1** output signals, we have added a "filter" process[30] consuming the chemical product $NAD^+$ and converting it back to NADH for small input concentrations (Scheme 1). This has allowed us to achieve high-quality signal separation for times as large as 600 sec and beyond, as described below.

We note that such processes can likely be implemented for any so-called $NAD^+$-dependent dehydrogenase,[33,34] e.g., glucose dehydrogenase activated by physiological amounts of glucose. However, aiming at the ultimate application of our system in physiological environment, we selected G6PDH, activated by Glc6P, which does not interfere with glucose naturally existing in blood. The filter system works in the following way: In the presence of G6PDH and Glc6P, the biocatalytically produced $NAD^+$ is converted back to NADH, thus preventing the absorbance changes, until Glc6P is totally consumed, then $NAD^+$ starts to accumulate resulting in the absorbance decrease. The delay in the biocatalytic formation of $NAD^+$ is controlled by the amount of Glc6P added to the system and should be optimized. Comprehensive approach to the filter performance optimization could include detailed analysis of the reactions kinetics.[30] However, a simple experimental optimization applied in the present study might suffice.

Application of the G6PDH-Glc6P filter following the biocatalytic cascade activated by ALT-LDH biomarker inputs (Scheme 1), has allowed a much better separation of the output signals generated by the system for the **1**,**1** vs. all the other combinations of the inputs (Figure 1B). However, while improving the binary-signal separation, such filtering can decrease the overall signal strength which could be an added source of relative noise.[28] Thus, this approach is useful for larger times when the decrease in the absorbance reaches its saturation, of relevance in actuation applications.[32] When the output signals were measured at 600 sec, the desired system operation corresponding to high-tolerance **AND**-logic realization was obtained in the presence of the filter (Figure 1B, Inset: red bars). Good-quality separation of the **0** and **1** output signals was found to persist at much larger times as well, up to 3 hours. The robustness of this analytical system has allowed its practical use in human serum solutions (see Supporting Information).

Receiver operating characteristic (ROC) analysis[35] was used to evaluate the performance of the LI diagnostic test with and without the added signal filtering process (see Supporting Information for details). The levels of biomarkers were chosen according to the published physiological and pathophysiological concentrations relevant for the diagnosis of the liver injury.[31] The present ROC analysis assesses the diagnostic properties due only to the analytical variance. Signal distributions in real medical-application situations can be different: additional comments on the "digital," binary, and analog-to-digital conversion nature of the information processing here are offered in the Supporting Information. The reaction time at which signal values were measured was 600 sec. The area under the ROC curve (AUC) is a single measure summarizing the overall accuracy of a test. It represents the probability that the diagnostic test will correctly distinguish between the physiological and the pathophysiological samples.[35] The AUC from empirical and smooth ROC curves,[36] which expectedly give consistent results in this case (and the corresponding 95% confidence intervals; CI) were estimated for LI diagnostic system. The AUC without a filter, from the empirical ROC curve was 0.92 (95% CI: 0.79-1.00), and 0.90 (95% CI: 0.77-1.00) from the smooth ROC curve (Figure 2). On the other hand, the diagnostic system with enabled filter has no overlap between output signal values **0** and **1**, i.e., of the ability to determine the output signals **0** and **1**, it offers the so-called "perfect performance"[35] (of course, only as far as analytical-procedure variance goes) of a diagnostic test. The ROC curve has AUC of 1.00 (95% CI: 1.00-1.00), a result corresponding to 100% sensitivity and 100% specificity. For details see Supporting Information, where similar results of ROC analysis are presented demonstrating an improved operation of the STI and ABT systems with the new filter process.

**CONCLUSIONS**

Analytical systems performing binary **AND** logic operations for the analysis of biomarkers characteristic of LI, STI and ABT were improved by integrating them with the filter process which converts the output signal back to the original chemical up to the certain optimized extend. The usefulness of the added filtering for a reliable operation of "digital-logic" biochemical analytical systems was established. The next step in the development of complex information and signal processing systems for analysis of injury biomarkers will involve analysis of samples from injured animals and human patients. This work is currently in progress and will be reported elsewhere.

**ASSOCIATED CONTENT**

**Supporting Information.** Composition and results for logic systems for the analysis of STI and ABT operated in buffer solutions, as well as for the LI system operated in human serum, are presented. This material is available free of charge via the Internet at http://pubs.acs.org.

**AUTHOR INFORMATION**

* Corresponding authors:
Tel: +1 (315) 268 4421; Fax: +1 (315) 268 6610; E-mail: ekatz@clarkson.edu (E. Katz)
Tel: +1 (858) 246 0128; Fax: +1 (858) 534 9553; E-mail: josephwang@ucsd.edu (J. Wang)

**ACKNOWLEDGMENT**

This work was supported by the Office of Naval Research (Award #N00014-08-1-1202) and NSF (Award # CBET-1066397).

# REFERENCES

(1) De Silva, A. P.; Uchiyama, S.; Vance, T. P.; Wannalerse, B. *Coord. Chem. Rev.* **2007**, *251*, 1623–1632.

(2) Szacilowski, K. *Chem. Rev.* **2008**, *108*, 3481–3548.

(3) Credi, A. *Angew. Chem. Int. Ed.* **2007**, *46*, 5472–5475.

(4) Pischel, U. *Angew. Chem. Int. Ed.* **2007**, *46*, 4026–4040.

(5) Andreasson, J.; Pischel, U. *Chem. Soc. Rev.* **2010**, *39*, 174–188.

(6) Shapiro, E.; Gil, B. *Nat. Nanotechnol.* **2007**, *2*, 84-85.

(7) Benenson, Y. *Mol. Biosyst.* **2009**, *5*, 675-685.

(8) Ashkenasy, G.; Ghadiri, M. R. *J. Am. Chem. Soc.* **2004**, *126*, 11140-11141.

(9) Stojanovic, M. N.; Stefanovic, D.; LaBean, T.; Yan, H. In: *Bioelectronics: From Theory to Applications*, Willner, I.; Katz, E. (Eds.) Wiley-VCH, Weinheim, **2005**, pp. 427-455.

(10) Katz, E.; Privman, V. *Chem. Soc. Rev.* **2010**, *39*, 1835-1857.

(11) Unger, R.; Moult, J. *Proteins* **2006**, *63*, 53-64.

(12) Ezziane, Z. *Nanotechnology*, **2006**, *17*, R27-R39.

(13) Rinaudo, K.; Bleris, L.; Maddamsetti, R.; Subramanian, S.; Weiss, R.; Benenson, Y. *Nat. Biotechnol.* **2007**, *25*, 795-801.

(14) Tamsir, A.; Tabor, J. J.; Voigt, C. A. *Nature* **2011**, *469*, 212-215.

(15) Li, Z.; Rosenbaum, M. A.; Venkataraman, A.; Tam, T. K.; Katz, E.; Angenent L. T. *Chem. Commun.* **2011**, *47*, 3060-3062.

(16) Kahan, M.; Gil, B.; Adar, R.; Shapiro, E. *Physica D*, **2008**, *237*, 1165-1172.

(17) Wang, J.; Katz, E. *Anal. Bioanal. Chem.* **2010**, *398*, 1591-1603.

(18) Adar, R.; Benenson, Y.; Linshiz, G.; Rosner, A.; Tishby, N.; Shapiro, E. *Proc. Natl. Acad. USA* **2004**, *101*, 9960-9965.

(19) May, E. E.; Dolan, P. L.; Crozier, P. S.; Brozik, S.; Manginell, M. *IEEE Sensors Journal* **2008**, *8*, 1011-1019.

(20) von Maltzahn, G.; Harris, T. J.; Park, J.-H.; Min, D.-H.; Schmidt, A. J.; Sailor, M. J.; Bhatia, S. N. *J. Am. Chem. Soc.* **2007**, *129*, 6064-6065.

(21) Zhou, J.; Halámek, J.; Bocharova, V.; Wang, J.; Katz, E. *Talanta* **2011**, *83*, 955-959.

(22) Halámek, J.; Bocharova, V.; Chinnapareddy, S.; Windmiller, J.R.; Strack, G.; Chuang, M.-C.; Zhou, J.; Santhosh, P.; Ramirez, G. V.; Arugula, M. A.; Wang, J.; Katz, E. *Mol. Biosyst.* **2010**, *6*, 2554-2560.

(23) Halámek, J.; Windmiller, J. R.; Zhou, J.; Chuang, M.-C.; Santhosh, P.; Strack, G.; Arugula, M. A.; Chinnapareddy, S.; Bocharova, V.; Wang, J.; Katz, E. *Analyst* **2010**, *135*, 2249-2259.

(24) Windmiller, J. R.; Strack, G.; Chuang, M.-C.; Halámek, J.; Santhosh, P.; Bocharova, V.; Zhou, J.; Katz, E.; Wang, J. *Sens. Actuat. B* **2010**, *150*, 285-290.

(25) Manesh, K. M.; Halámek, J.; Pita, M.; Zhou, J.; Tam, T. K.; Santhosh, P.; Chuang, M.-C.; Windmiller, J. R.; Abidin, D.; Katz, E.; Wang, J. *Biosens. Bioelectron.* **2009**, *24*, 3569-3574.

(26) Melnikov, D.; Strack, G.; Zhou, J.; Windmiller, J. R.; Halámek, J.; Bocharova, V.; Chuang, M.-C.; Santhosh, P.; Privman, V.; Wang, J.; Katz, E. *J. Phys. Chem. B* **2010**, *114*, 12166-12174.

(27) Tokarev, I.; Gopishetty, V.; Zhou, J.; Pita, M.; Motornov, M.; Katz, E.; Minko, S. *ACS Appl. Mater. Interfaces* **2009**, *1*, 532-536.

(28) Privman, V. *Isr. J. Chem.* **2011**, *51*, 118-131.

(29) Pita, M.; Privman, V.; Arugula, M. A.; Melnikov, D.; Bocharova, V.; Katz, E. *PhysChemChemPhys* **2011**, *13*, 4507-4513.

(30) Privman, V.; Halámek, J.; Arugula, M. A.; Melnikov, D.; Bocharova, V.; Katz, E. *J. Phys. Chem. B* **2010**, *114*, 14103-14109.

(31) Tan, K.-K.; Bang, S.-L.; Vijayan, A.; Chiu, M.-T. *Injury* **2009**, *40*, 978-983.

(32) Privman, M.; Tam, T. K.; Bocharova, V.; Halámek, J.; Wang, J.; Katz, E. *ACS Appl. Mater. Interfaces* **2011**, *3*, 1620-1623.

(33) Rosemeyer, M. A. *Cell Biochem. Funct.* **1987**, 5, 79-95.

(34) Beutler, E. *Blood* **2008**, *111*, 16-24.

(35) Zweig, M. H.; Campbell, G. *Clin. Chem.*, **1993**, *39*, 561-577.

(36) Faraggi, D.; Raiser, B. *Stat. Med.* **2002**, *21*, 3093-3106.

**SCHEME AND FIGURE CAPTIONS**

**Scheme 1.** The biocatalytic cascade operating as **AND** logic gate for analysis of LI, (A) without, and (B) with the added biocatalytic-filter reaction. The following abreviations for products and intermediates are used: Pyr for pyruvate, Lac for lactate, Glu for glutamate and 6-PGluc for 6-phospho-gluconic acid. Other chemicals and the operation of the system are detailed in the Experimental section.

**Figure 1.** Time dependent absorbance changes generated by the system outlined in Scheme 1, (A) without, and (B) with the added biocatalytic-filter reaction upon application of various combinations of the ALT and LDH input signals. The Inset shows the output signals measured at 600 sec, obtained without and with the filter process, for different combinations of the inputs. The vertical axis in the inset gives the normalized values of the averaged absorbance decrease, $\Delta A_n = \Delta A / \Delta A_{max}$, ranging between 0 and 1, where $A_{max}$ is the experimental mean value for the logic **1** output (the error bars shown are also normalized).

**Figure 2.** Receiver operating characteristic (ROC) empirical (green) and smoothed (blue) curves for the non-filtered system for the LI. The two (coinciding) ROC curves for the filtered system are shown in red. Random choice is denoted by the gray diagonal line; the best cutoffs for calculating the maximized-accuracy detection are indicated as the color-coded symbols (for details see Supporting Information).

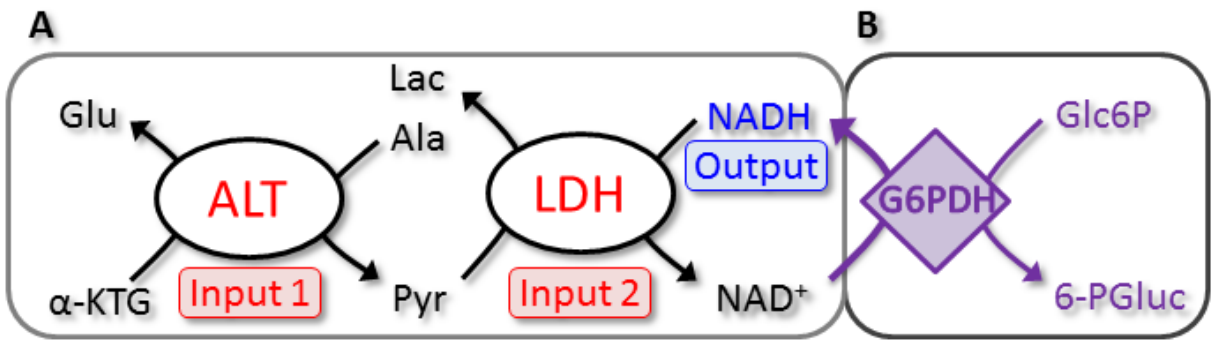

**Scheme 1.**

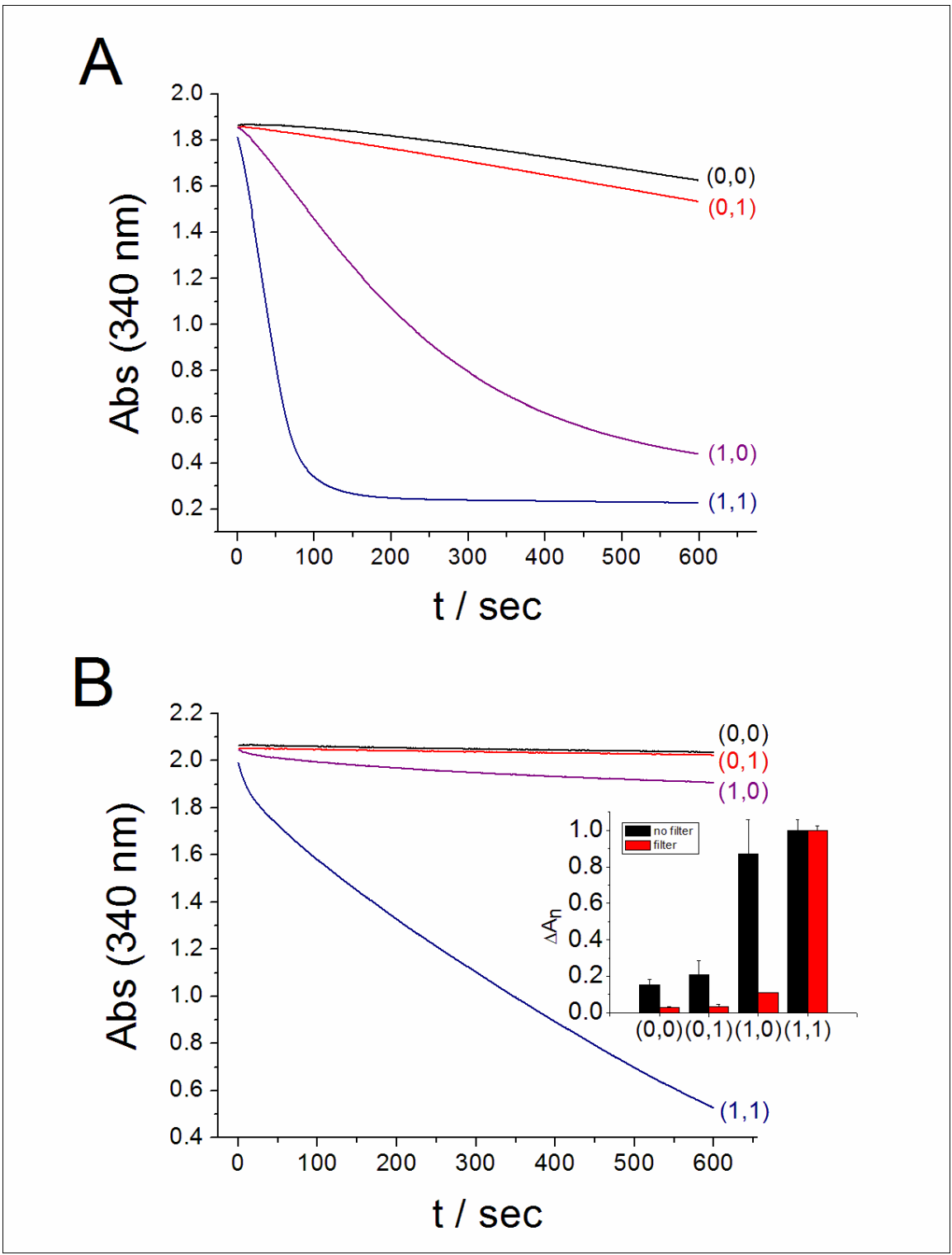

**Figure 1.**

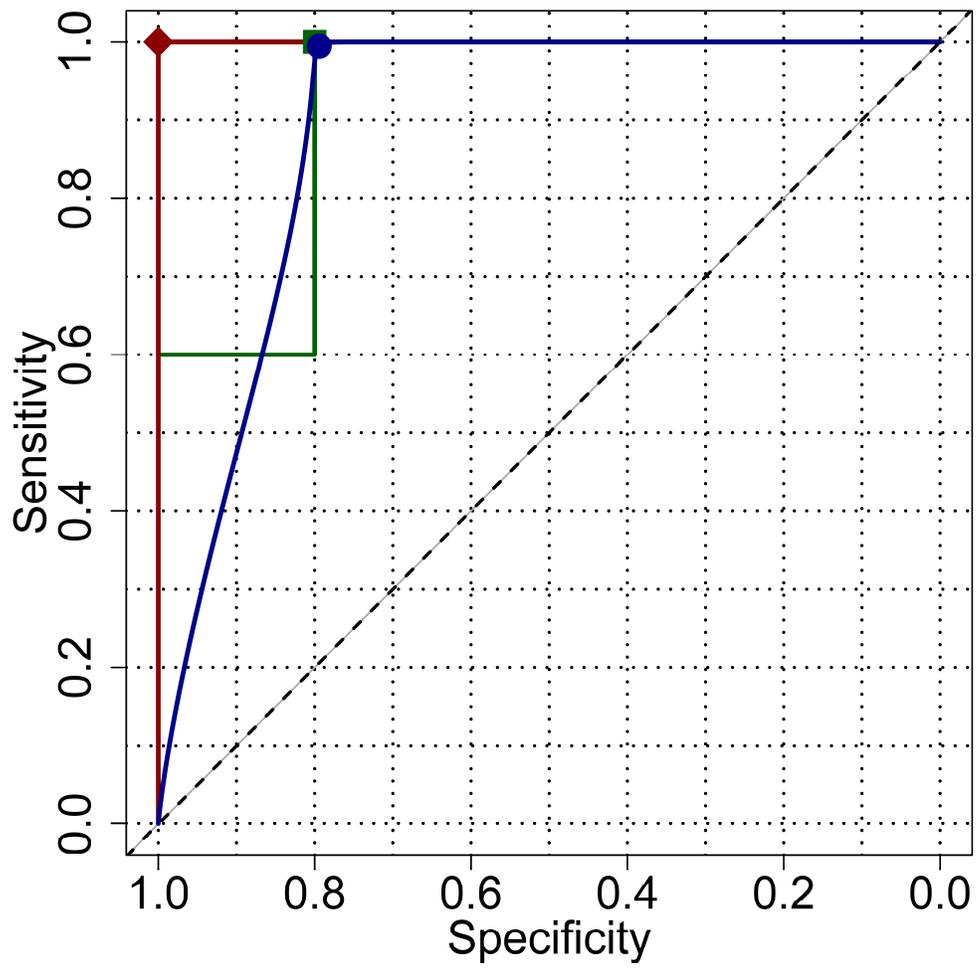

Figure 2.

**For TOC only**

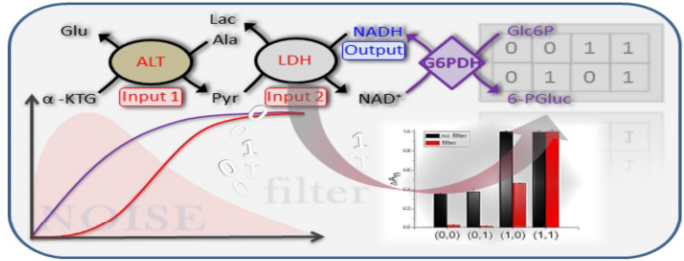

# Biomolecular Filters for Improved Separation of Output Signals in Enzyme Logic Systems Applied to Biomedical Analysis

Jan Halámek,[a] Jian Zhou,[a] Lenka Halámková,[a,c] Vera Bocharova,[a] Vladimir Privman,[b] Joseph Wang[d]* and Evgeny Katz[a]*

[a] Department of Chemistry and Biomolecular Science,  [b] Department of Physics,

[c] Department of Biology, Clarkson University, Potsdam, NY 13699, USA

[d] Department of NanoEngineering, University of California at San Diego, La Jolla, CA 92093, USA

---

* Corresponding authors:

Tel: +1 (315) 268 4421; Fax: +1 (315) 268 6610; E-mail: ekatz@clarkson.edu (E. Katz)

Tel: +1 (858) 246 0128; Fax: +1 (858) 534 9553; E-mail: josephwang@ucsd.edu (J. Wang)

## *Supporting Information*

**Summary of experimental results and their statistical treatment for the STI and ABT systems operating in buffer solutions and for the LI system operating in human serum solution**

**Experimental Section**

*Chemicals and materials:* Alanine transaminase from porcine heart (ALT, E.C. 2.6.1.2), glucose-6-phosphate dehydrogenase from *Leuconostoc mesenteroides* (G6PDH, E.C. 1.1.1.49), microperoxidase-11 (MP-11), lactate dehydrogenase from porcine heart (LDH, E.C. 1.1.1.27), pyruvate kinase from rabbit muscle (PK, E.C. 2.7.1.40), creatine kinase from rabbit muscle (CK, E.C. 2.7.3.2), glycyl-glycine (Gly-Gly), *tris*(hydroxymethyl) aminomethane hydrochloride (Tris-buffer), L-alanine (Ala), α-ketoglutaric acid (α-KTG), β-nicotinamide adenine dinucleotide reduced dipotassium salt (NADH), β-nicotinamide adenine dinucleotide dipotassium salt (NAD$^+$), L(+)-lactic acid (Lac), D-glucose 6-phosphate sodium salt (Glc6P), creatine anhydrous (Crt), phospho(enol)pyruvate monopotassium salt (PEP), adenosine 5'-triphosphate disodium salt (ATP, from bacterial source), magnesium acetate tetrahydrate (MgAc$_2$), potassium hydroxide (KOH), hydrochloric acid (HCl) and serum from human male AB plasma were purchased from Sigma-Aldrich and were used as supplied without further purification or pretreatment. Hydrogen peroxide (H$_2$O$_2$) (30% w/w) was purchased from Baker. Ultrapure deionized water (18.2 MΩ·cm) from a NANOpure Diamond (Barnstead) source was used in all of the experiments.

*Instrumentation and measurements:* In order to mimic physiological conditions, optical measurements were done in temperature-controlled 1 mL poly(methyl methacrylate) (PMMA) cuvettes at $37.0 \pm 0.2$°C with 1 cm pathway using Shimadzu UV-2450 UV-Vis spectrophotometer (with a TCC-240A temperature-controlled holder). All reagents were incubated at this temperature prior to experimentation. A Mettler Toledo SevenEasy s20 pH-meter was employed for the pH measurements.

**Composition and operation of systems for the analysis of injuries**

*LI system (serum experiment):* Human serum was diluted to 50% by the Tris-buffer, pH 7.4. The final concentrations of the logic system "machinery" and components of the filter were Ala (200 mM), α-KTG (10 mM), NADH (150 μM), Glc6P (4 mM) and G6PDH (10 U·mL$^{-1}$). ALT and LDH used as biomarkers of liver injury were dissolved in pure human serum. Because of dilution, logic **0** and **1** levels of ALT (0.01 U·mL$^{-1}$ and 1 U·mL$^{-1}$) and LDH (0.075 and 0.5 U·mL$^{-1}$) were used as half of the physiological and pathophysiological values.[1] Input signals were applied to the logic system in order to realize meaningful circulating levels of these biomarkers and perform the **AND** logic operation with filter. The output signal corresponding to the decreasing concentration of NADH was measured optically at λ = 340 nm.

*STI system (buffer experiment):* Gly-Gly buffer, 50 mM, with MgAc$_2$ (6.7 mM) was titrated with KOH to the pH value of 7.95 and used as a background solution (note that Mg$^{2+}$ and K$^+$ cations are essential for activation of CK and PK, respectively). The following components were dissolved in this solution to perform the **AND** logic operation: NADH (0.1 mM), ATP (1 mM),

PEP (1.5 mM), PK (1.6 U·mL$^{-1}$), Crt (7.5 mM). The filtering compounds Glc6P (0.1 mM) and G6PDH (2 U·mL$^{-1}$) were prepared in the same buffer. Logic **0** and **1** levels of CK (0.1 and 0.71 U·mL$^{-1}$) and LDH (0.15 and 1 U·mL$^{-1}$) input signals were applied to the logic system in order to realize meaningful circulating levels of these biomarkers.[2,3] The output signal corresponding to the decreasing concentration of NADH was measured optically at $\lambda$= 340 nm.

*ABT system (buffer experiment):* Gly-Gly buffer, 50 mM, pH 8.5 tuned by KOH, containing MgAc$_2$ (6.7 mM) and NAD$^+$ (10 mM) was used to perform the **AND** logic operation. MP-11 (50 µM) and H$_2$O$_2$ (1.5 mM) were used as filter components. Logic **0** and **1** levels of LDH (0.15 and 1.0 U·mL$^{-1}$) and Lac (1.6 and 6.0 mM) input signals were applied to the logic system in order to realize meaningful circulating levels of these biomarkers.[3,4] The output signal corresponding to the NADH formation was measured optically at $\lambda$ = 340 nm.

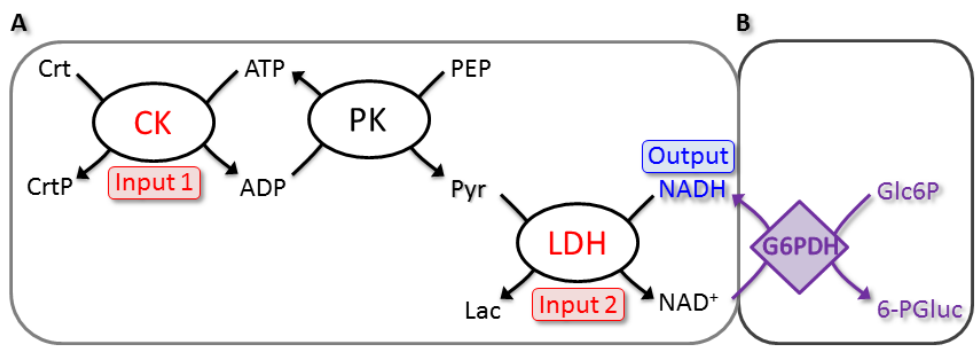

**Figure SI 1.** The biocatalytic cascade operating as the **AND** logic gate for analysis of the STI without (A) and with presence (B) of the biocatalytic filter. The following abreviations for products and intermediates are used: CrtP = phosphocreatine, ADP = adenosine diphosphate, Pyr = pyruvate, 6-PGluc = 6-phospho-gluconic acid. Other abbreviations are specified in the *Chemicals and materials* section.

**Results for the STI and ABT detection systems measured in buffer solutions and for the LI system measured in human serum solutions**

*Soft Tissue Injury (STI) system operating in buffer:* Two enzymes, CK and LDH, were applied as biomarkers characteristic of soft tissue injury.[2,3] Their simultaneous increase from normal to pathophysiological concentrations provides an evidence of STI conditions. The biochemical cascade catalyzed in the presence of the both enzyme-biomarkers (note the biocatalytic operation of PK being a part of the logic gate "machinery") results in the oxidation of NADH to $NAD^+$ (Figure SI 1A), thus yielding the corresponding absorbance decrease. The *absolute value of the absorbance change* was used to define the output signal produced by the system. The logic value of the output signal changes from the low **0** value to the high **1** value only upon the concerted work of the both enzyme-inputs (logic inputs combination **1,1**), thus mimicking the **AND** logic operation. Since the logic **0** values of the input signals are not physical zero concentrations (they rather correspond to the normal physiological concentrations of the enzymes), the NADH absorbance is also changing upon other combinations of the inputs (**0,0**; **0,1**; **1,0**). Similarly to the LI system described in the main text of the paper, the STI system operation was improved upon addition of the filter system to the analyzing biocatalytic cascade (Figure SI 1). The experimental data obtained in the presence and absence of the filter are summarized in Figure SI 2 and statistically analyzed in Figure SI 3, as further detailed below.

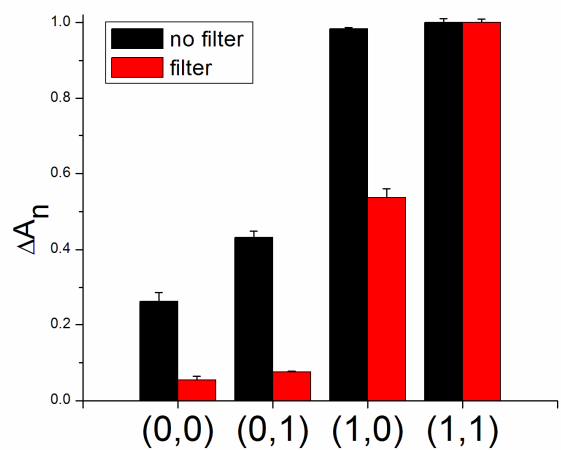

**Figure SI 2.** Bar chart featuring the **AND** logic operation of the optical system for detection of the STI. The black-colored bars indicate performance of the STI system without filter whereas the red-colored bars are with the applied filter. Optical absorbance measurements were performed at $\lambda = 340$ nm at time of 350 sec.

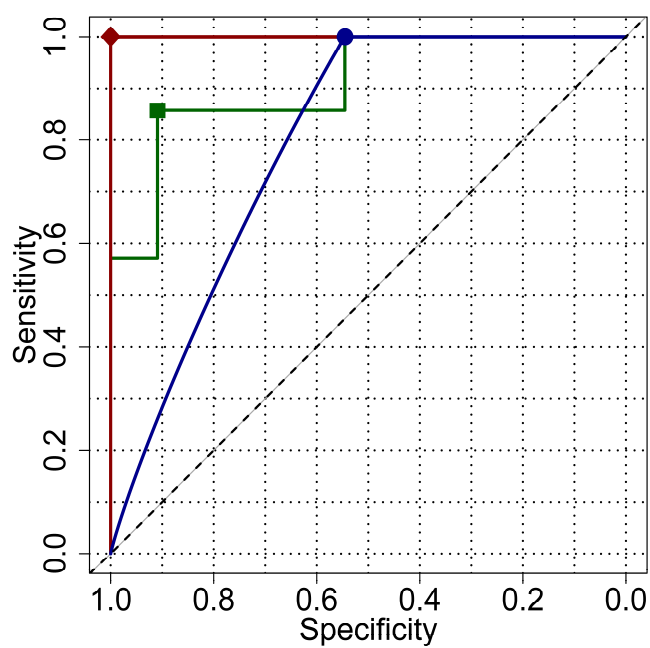

**Figure SI 3**. Receiver operating characteristic (ROC) empirical (green line) and smoothed (blue line) curve for the non-filtered system in detection of the STI. ROC curves for the filtered system are shown in red color. Note the both empirical and smooth ROC curves for the filtered system correspond to the "perfect performance." Random choice is denoted by the solid diagonal line. The best cutoffs which maximize the accuracy are indicated as solid symbols.

*Abdominal Trauma (ABT) system operating in buffer:* The enzyme LDH and its substrate Lac appearing together at elevated concentrations can be used as biomarkers of ABT.[3,4] The biocatalytic reaction activated by the enzyme and the corresponding substrate, results in the reduction of $NAD^+$ cofactor (Figure SI 4), thus leading to increased absorbance at $\lambda = 340$ nm corresponding to the formation of NADH. The *absorbance change* was defined as the output signal produced by the system. The logic value of the output signal changes from the low **0** value to the high **1** value only upon the concerted work of both inputs (logic inputs combination **1,1**), thus mimicking **AND** logic operation. Since the logic **0** values of the input signals are not physical zero concentrations (they correspond to the normal physiological concentrations of the enzyme and its substrate), the NADH absorbance is also changing upon other combinations of the inputs (**0,0**; **0,1**; **1,0**). Similarly to the LI system described in the main text of the paper, the ABT system operation was improved upon addition of the filter process to the analyzing biocatalytic cascade (Figure SI 4). The experimental data obtained in the presence and absence of the filter are summarized in Figure SI 5 and statistically analyzed in Figure SI 6.

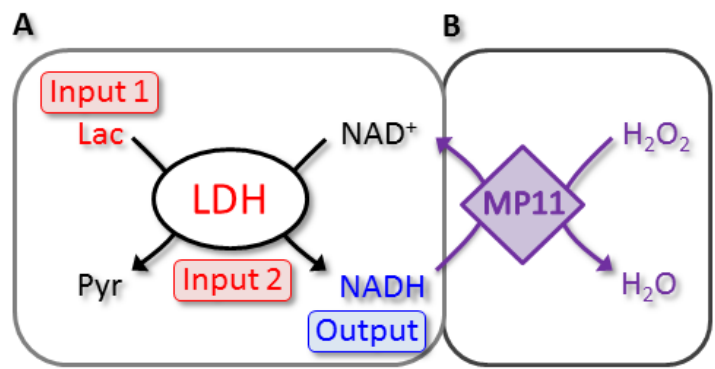

**Figure SI 4.** The biocatalytic cascade operating as an **AND** logic gate for analysis of the ABT in the absence (A) and in the presence (B) of the biocatalytic filter. The following abreviation for a product is used: Pyr = pyruvate. Other abbreviations are specified in the *Chemicals and materials* section.

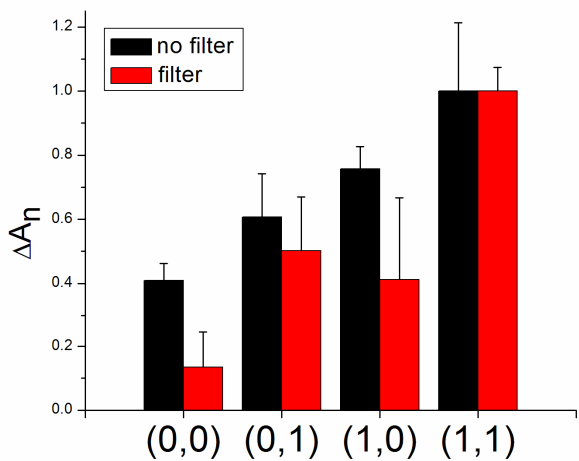

**Figure SI 5.** Bar chart featuring the **AND** logic operation of the optical system for detection of ABT. The corresponding combinations of input signals without a filter (the black bars) and with a filter (the red bars) are indicated. Optical absorbance measurements were performed at $\lambda$ = 340 nm at time of 1200 sec.

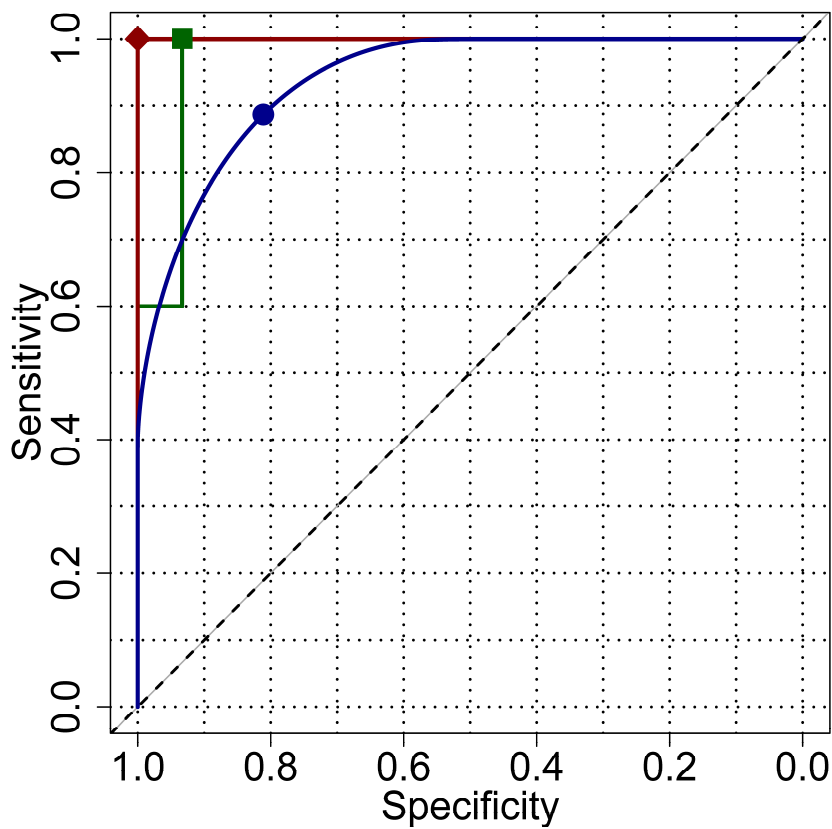

**Figure SI 6.** Receiver operating characteristic (ROC) empirical (green line) and smoothed (blue line) curve for non-filtered system in detection of the ABT. ROC curves for filtered system are shown in red color. Note the both empirical and smooth ROC curves for filtered system correspond to the "perfect performance." Random choice is denoted by the grey diagonal line. The best cutoffs which maximize the accuracy are indicated as solid symbols.

*Liver Injury (LI) system operating in serum:* The system is the same as described in the main part of the paper (Scheme 1; Figure SI 7), but its operation was studied in the presence of human serum solution. The experimental data obtained in the presence and absence of the filter are summarized in Figure SI 8 and statistically analyzed in Figure SI 9.

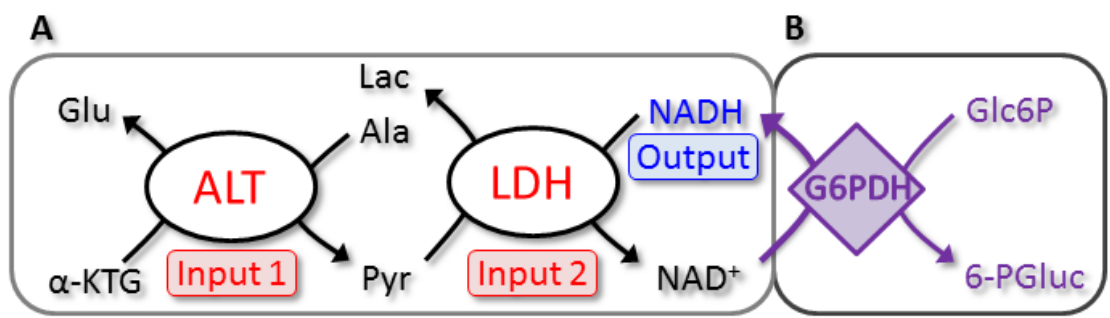

**Figure SI 7** (*the same as Scheme 1 in the paper*). The biocatalytic cascade operating as the **AND** logic gate for analysis of LI in the absence (A) and in the presence (B) of the biocatalytic filter. The following abreviation for a product is used: 6-PGluc = 6-phosphogluconic acid, Pyr = pyruvate, Lac = lactate, Glu = glutamate. Other abbreviations are specified in the *Chemicals and materials* section.

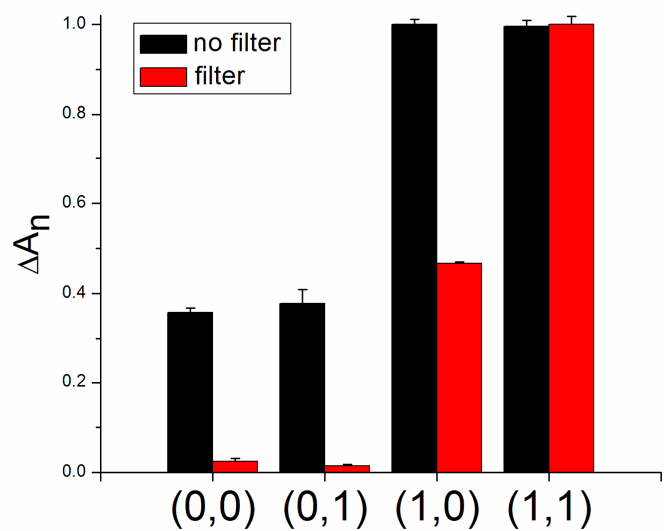

**Figure SI 8.** Bar charts featuring the **AND** logic operation of the optical system for detection of LI. The corresponding combinations of input signals without a filter (the black bars) and with a filter (the red bars) are indicated.

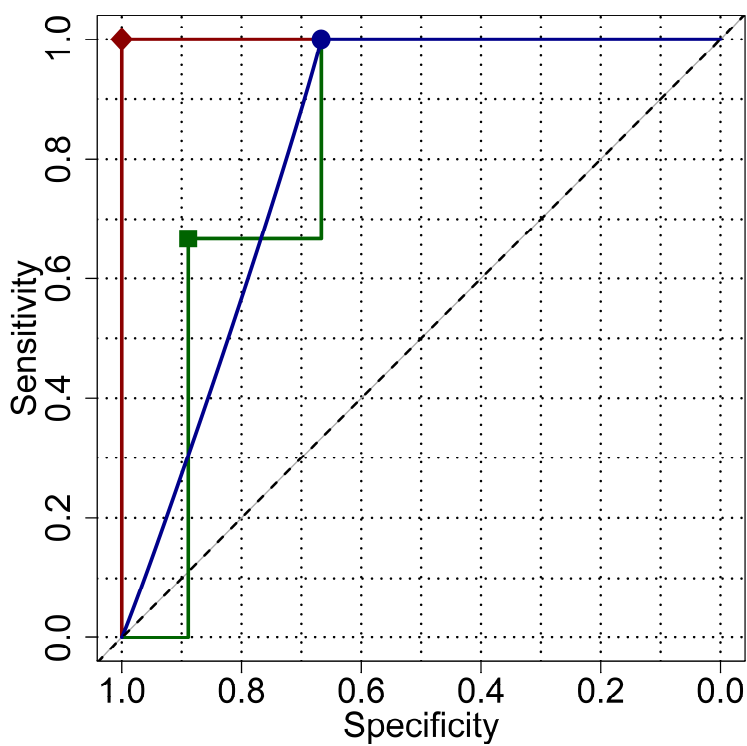

**Figure SI 9.** Receiver operating characteristic (ROC) empirical (green line) and smoothed (blue line) curve for non-filtered system in detection of LI in serum. ROC curves for filtered system are shown in red color. Note the both empirical and smooth ROC curves for filtered system correspond to the "perfect performance." Random choice is denoted by the grey diagonal line. The best cutoffs which maximize the accuracy are indicated as solid symbols.

**Statistical data analysis**

The biochemical cascade composed of the specific biocatalytic reactions, results in a distinct change in the output signal (change in absorbance) only in case of the cooperative action of both biomarkers (logic input combination **1**,**1**) and logic value of the output attains **1**. Other combinations of the input signals (**0**,**0**, **0**,**1** and **1**,**0**) represent logic value of the output **0**, but this occurs for non-zero physical values of the inputs, due to their physiological nature. The effect of the enzyme filter on improving the discrimination of logic output **0** and **1** was evaluated by Receiver Operating Characteristic (ROC) curves.

The accuracy of the diagnostic method depends on the ability to distinguish between pathophysiological and physiological samples being tested, which represent the logic output **1** and **0**, respectively. The "area under curve" (AUC) is a summary single measure, defined as an area under an ROC curve, which combines concepts of sensitivity and specificity and is commonly used for quantification of diagnostic test accuracy. The sensitivity—the "true positive rate" (TPR), and specificity—the "true negative rate" (TNR), both depend on the tested thresholds; the TPR and TNR vary as the threshold varies. By considering various possible values of the threshold, an ROC curve can be constructed as a continuous function of sensitivity versus specificity (possibly 1-specificity—the "false positive rate"; FPR).[5] AUC of 1 represents "perfect performance," i.e. 100% TPR, at TNR of 100%; AUC of 0.5 represents a random diagnosis.

The AUCs of empirical ROC curves were estimated by the trapezoidal method of integration, and the corresponding 95% confidence intervals (CI) were estimated with the method described by De Long et al.[6] Using the ROC analysis, the best thresholds (above which the absorbance change is considered as a logic output **1**) that yielded the maximum accuracy were determined and characterized by their specificity and sensitivity with the corresponding 95% CIs. Smoothed ROC curves were additionally estimated by using a non-parametric method. The Kernel density function[5] was used to fit smooth ROC curves to data points because this method is free of parametric assumptions.[7] This smoothed-curve method outperforms the competing methods when pathophysiological and/or control group has a bimodal distribution (see the differences in absorbance changes between logic inputs **0**,**0**, **0**,**1** and logic input **1**,**0**). The bandwidth of the Kernel function is fixed using the robust method developed by Sheather and Jones.[8] The AUCs of smooth ROC curves are indicated with corresponding 95% CIs computed with 2000 stratified bootstrap replicates as described elsewhere.[9] All statistic tests and data plotting were performed using the standard R-project software R 2.1, available online[10].

Three enzymatic systems described above were evaluated by ROC curve analysis before and after the enzymatic filter was included. The cutoff values, sensitivity, specificity, and area under the empirical and smooth ROC curve for all the enzymatic systems without and with filter are presented in Table 1. In non-filtered enzymatic systems, the ability of the systems to distinguish the logic output **0** from **1** was found as follows: (*i*) liver injury measured in serum (AUC 0.82, 95% CI 0.55-1.00), (*ii*) soft tissue injury measured in buffer (AUC 0.91, 95% CI 0.76-1.00), and (*iii*) abdominal trauma measured in buffer (AUC 0.97, 95% CI 0.91-1.00). The AUCs for smooth ROC curves for non-filtered enzymatic systems were as follows: (*i*) liver injury measured in serum (AUC 0.83, 95% CI 0.65-1.00), (*ii*) soft tissue injury measured in buffer (AUC 0.80, 95% CI 0.73-0.95) and (*iii*) abdominal trauma measured in buffer (AUC 0.94, 95% CI 0.87-1.00). Using the enzymatic filter, we have achieved "perfect performance" in terms of the AUC for both empirical and smooth ROC curve in all diagnostic systems. ROC curves showed a very good discrimination between logical output **0** and **1**, with an AUC of 1.00 (95% CI 1.00-1.00) for all filtered systems; such values correspond with 100% sensitivity and 100% specificity. Notice that in the case of the "perfect performance" the AUCs of empirical and smooth ROCs curve are the same.

**Table 1**: Receiver Operating Characteristic Curve Analysis of the enzymatic systems for diagnosis of liver injury (performed in buffer and serum), soft tissue injury and abdominal trauma with the enabled/disabled biocatalytic filter.

|  |  | AUC 95% CI | Cutoff | Sensitivity 95% CI | Specificity 95% CI | AUC (smooth ROC curve) 95% CI |
|---|---|---|---|---|---|---|
| LI (Buffer) | No Filter | 0.92 (0.79-1.00) | 1.70 | 0.60 (0.23-0.88) | 1.00 (0.80-1.00) | 0.90 (0.77-1.00) |
|  | Filter | 1.00 (1.00-1.00) | 0.84 | 1.00 (1.00-1.00) | 1.00 (1.00-1.00) | 1.00 (1.00-1.00) |
| LI (Serum) | No Filter | 0.82 (0.55-1.00) | 0.77 | 0.67 (0.21-0.94) | 0.89 (0.57-0.98) | 0.83 (0.65-1.00) |
|  | Filter | 1.00 (1.00-1.00) | 0.35 | 1.00 (1.00-1.00) | 1.00 (1.00-1.00) | 1.00 (1.00-1.00) |
| STI (Buffer) | No Filter | 0.91 (0.76-1.00) | 0.58 | 0.86 (0.49-0.97) | 0.91 (0.62-0.98) | 0.80 (0.73-0.95) |
|  | Filter | 1.00 (1.00-1.00) | 0.47 | 1.00 (1.00-1.00) | 1.00 (1.00-1.00) | 1.00 (1.00-1.00) |
| ABT (Buffer) | No filter | 0.97 (0.91-1.00) | 0.76 | 1.00 (0.57-1.00) | 0.93 (0.70-0.99) | 0.94 (0.87-1.00) |
|  | Filter | 1.00 (1.00-1.00) | 0.48 | 1.00 (1.00-1.00) | 1.00 (1.00-1.00) | 1.00 (1.00-1.00) |

Finally, as stressed in the main text, we emphasize that, our conclusion of "perfect performance" applies only to the analytical variance of the process of the binary **AND** function generation and signal-conversion to digital answers. Indeed, we do not probe the additional noise effects due to actual clinical-testing concentration distributions of the biomarkers involved. The reason has been that the details of the latter distributions[11,12] are simply not well known presently, even though all the considered biomarkers are used in actual biomedical testing, in different assay formats[13,14] than those proposed here. As a precaution, we sat our **0** and **1** "digital"—perhaps more carefully termed "binary"—values safely at the edges of the approximately known[2-4,15] physiological and pathophysiological ranges: **0** approximately at the highest value of the lower range, **1** approximately at the lowest value of the upper range. In summary, our actual information processing thus involves the binary **AND** function accompanied by analog-to-digital signal conversion for the output, realized and made high-fidelity of adding the "filtering process" to the enzymatic reaction cascade.

## ACKNOWLEDGMENT

This work was supported by the Office of Naval Research (Award #N00014-08-1-1202) and NSF (Award # CBET-1066397).

## REFERENCES

1. Tan, K.-K.; Bang, S.-L.; Vijayan, A.; Chiu, M.-T., *Injury* **2009**, *40*, 978-983.
2. Olerud, J. E.; Homer, L. D.; Carroll, H. W. *Arch. Int. Med.* **1976**, *136*, 692-697.
3. Kratz, A.; Ferraro, M.; Sluss, P. M.; Lewandrowski, K. B. *New Engl. J. Med.* **2004**, *351*, 1548-1563.
4. Sprules, S. D.; Hart, J. P.; Wring, S. A.; Pittson, R. *Anal. Chim. Acta* **1995**, *304*, 17-24.
5. Zou, K. H.; Hall, W. J.; Shapiro, D. E. *Stat. Med.* **1997**, *16*, 2143-2156.
6. DeLong, E. R.; DeLong, D. M.; Clarke-Pearson, D. L. *Biometrics* **1988**, *44*, 837-845.
7. Faraggi, D.; Raiser, B. *Stat. Med.* **2002**, *21*, 3093-3106.
8. Sheather, S. J.; Jones, M. C.; *J. Roy. Stat. Soc. B* **1991**, *53*, 683-690.

9. Carpenter, J.; Bithell, J. *Stat. Med.* **2000**, *19*, 1141-1164.

10. R-project, http://www.www.r-project.org

11. Bathum, L.; Petersen, H. C.; Rosholm, J. U.; Hyltoft Petersen, P.; Vaupel, J; Christensen, K. *Clin. Chem.* **2001**, *47*, 81-87.

12. Pratt, D. S.; Kaplan, M. M. *N. Engl. J. Med.* **2000**, *342*, 1266-1271.

13. Henry, J. B. *Clinical Diagnosis and Management by Laboratory Methods*. 20th ed. Philadelphia: W. B. Saunders, **2001**.

14. Wallach, J. *Interpretation of Diagnostic Tests*. 7th ed. Philadelphia: Lippincott Williams & Wilkins, **2000**.

15. Tan, K.-K.; Bang, S.-L.; Vijayan, A.; Chiu, M.-T. *Injury* **2009**, *40*, 978-983.